# Analysis and Comparison of Different Wavelet Transform Methods Using Benchmarks for Image Fusion

. [A]G. Karpaga Kannan, [B]T. Deepika.

[A]Assistant Professor, [B]PG Student.

[A,B,] RVS College of Engineering Technology, Dindigul, Tamilnadu

[A]karpagakannan@gmail.com, [B]deepikarvs@gmail.com.

**Abstract-** In recent years, many research achievements are made in the medical image fusion field. Medical Image fusion means that several of various modality image information are comprehended together to form one image to express its information. The aim of image fusion is to integrate complementary and redundant information. CT/MRI is the one of the most common medical image fusion. These medical modalities give information about different diseases. Complementary information is offered by CT and MRI. CT provides best information about denser tissue and MRI offers better information on soft tissue. There are two approaches to image fusion, namely Spatial Fusion and Transform fusion. Transform fusion uses transform for representing the source images at multi scale. This paper presents a Wavelet Transform image fusion methodology based on the intensity magnitudes of the wavelet coefficients and compares five variations of the wavelet transform implemented separately in this fusion model. The image fusion model, using the Discrete Wavelet Transform (DWT), the Stationary Wavelet Transform (SWT), the Integer Lifting Wavelet Transform (ILFT) the dual tree Complex Wavelet Transform (DT CWT) and dual tree Q-shift dual tree CWT, is applied to multi-modal images. The resulting fused images are compared visually and through benchmarks such as Entropy (E), Peak Signal to Noise Ratio, (PSNR), Root Mean Square Error (RMSE), Image Quality Index (IQI) and Standard deviation (SD) computations.

## I. INTRODUCTION

The development of Fourier began with Joseph Fourier (1807) with his theories of frequency analysis, referred to as Fourier synthesis. After 1807, by exploring the meaning of functions, Fourier series convergence, and orthogonal systems, mathematicians gradually were led from their previous notion of *frequency analysis* to the notion of *scale analysis.* [1] Scale analysis is less sensitive to noise because it measures the average fluctuations of the signal at different scales. In time-frequency analysis of a signal, the classical Fourier transform analysis is inadequate because Fourier transform of a signal does not contain any local information. Fourier transform is a powerful tool for analyzing the components of a stationary signal. But it is failed for analyzing the non stationary signal where as wavelet transform allows the components of a non-stationary signal to be analyzed. To overcome this drawbacks, Dennis Gabor in 1946, first introduced the windowed-Fourier transform, i.e. short-time Fourier transform known later as Gabor transform. The windowed Fourier transform (WFT) is one solution to the problem of better representing the nonperiodic signal. The WFT can be used to give information about signals simultaneously in the time domain and in the frequency domain. To approximate a function by samples, and to approximate the Fourier integral by the discrete Fourier transform, requires applying a matrix whose order is the number sample points *n*. The Fourier matrix can be factored into a product of just a few sparse matrices, and the resulting factors can be applied to a vector in a total of order *n* log *n* arithmetic operations. This is the so-called *fast Fourier transform* or FFT. It is an efficient algorithm to compute discrete Fourier transform and inverse discrete Fourier transform. Due to the disadvantage of Fourier transform, which include localized only in frequency domain, not windows vary and lack of capability, approaches based on Wavelet transform have begun. Haar's started his work in the early 20th century. First wavelet discovered by Alfered Haar (1909). Signal transmission is based on transmission of a series of numbers. The series representation of a function is important in all types of signal transmission. The wavelet representation of a function is a new technique. Meyer [7] found the existing literature of wavelets. Later many eminent mathematicians e.g. I. Daubechies, A. Grossmann, S. Mallat, Y. Meyer, R. A. deVore, Coifman, V. Wickerhauser made a remarkable contribution to the wavelet theory. In 1982 Jean Morlet a French geophysicist, introduced the concept of a `wavelet'. The wavelet means small wave and the study of wavelet transform is a new tool for seismic signal analysis. Wavelets are well-suited for approximating data with sharp discontinuities. The wavelet analysis procedure is to adopt a wavelet prototype function, called an analyzing wavelet or mother wavelet. Mathematical formulation of signal expansion using wavelets gives wavelet transform pair, which is analogous to the Fourier transform pair. Wavelet transform of a function is the improved version of Fourier transform. Immediately, Alex Grossmann theoretical physicists studied inverse formula for the wavelet transform. The joint collaboration of Morlet and Grossmann [5] yielded a detailed mathematical study of the continuous wavelet transforms and their various applications, of course without the realization that similar results had already been obtained in 1950's by Calderon, Littlewood, Paley and Franklin. Wavelet Transform (WT) for representing the source image at multi scale. The most widely used transform for image fusion at multi scale is Discrete Wavelet Transform (DWT) since it minimizes structural distortions.DWT was invented by the Hungarian mathematician Alfred Haar. It was formulated by the Belgian mathematician Ingrid Daubechies in 1988. But, DWT suffers from lack of shift variance, aliasing, oscillations & poor directionality .One way to avoid these disadvantages is to use Complex Wavelet Transform. It is based on complex-valued oscillating sinusoids. CWT cannot exactly possess the analytic signal properties and will not perfectly overcome the four DWT shortcomings for that Kingsbury introduce Dual Tree complex wavelet transform (DTCWT), which is most expensive, computationally intensive, and approximately shift invariant [6-13]. But, the un-decimated DWT, namely Stationary Wavelet Transform (SWT) is shift invariant and Wavelet Packet Transform (WPT) provides more directionality. This benefit comes from the ability of the WPT to better represent high frequency content and high frequency oscillating signals in particular.
Wavelet transform verses Fourier transform

The fast Fourier transform (FFT) and the discrete wavelet transform (DWT) are both linear operations that generate a data structure that contains $\log_2 n$ segments of various lengths, usually filling and transforming it into a different data vector of length $2^n$. The inverse transform matrix for both the FFT and the DWT is the transpose of the original. Basic functions of both are localized in frequency. These are the similarities of WT and FT. The most interesting dissimilarity between these two kinds of transforms is that individual wavelet functions are localized in space. Fourier sine and cosine functions are not. Fourier transform does not give any information of the signal in the time domain. If a signal has a discontinuity, Fourier transform provides many coefficients with large magnitude. But WT generates a few significant coefficients around the discontinuity. The Fourier transform is less useful in analyzing non-stationary signal where as Wavelets also allow filters to be constructed for stationary and non-stationary signals Wells [4], Strang [2]. Wavelets often give a better signal representation using Multiresolution analysis Walnut [3]. Fourier transform is based on a single function $\psi(t)$ and that this function is scaled. But for the wavelet transform we can also shift the function, thus generating a two-parameter family of functions $\psi_{a,b}(t)$ defined by Debnath [5]. Wavelet theory is capable of revealing aspects of data that other signal analysis techniques miss the aspects like trends, breakdown points, and discontinuities in higher derivatives and self-similarity. The classical Fourier analysis is not suited for detecting them. It can often compress or de-noise a signal without appreciable degradation. Wavelet transform based approach is better technique and it takes less response time which is more suitable for online verification with high accuracy than that of Fourier transform technique.

## II. WAVELET TRANSFORM

Wavelet transform is an orthogonal transform, which is not only has the excellence of Fourier transform but also settle the contradiction in the spatial field and frequent field for the Fourier transform. There are two main groups of transforms, continuous and discrete. The wavelet representation then consists of the low-pass band at the lowest resolution and the high pass bands at each step. This transform is invertible and redundant. Wavelets as a family of functions constructed from translations and dilations of a Single function called the "mother wavelet" $\psi(t)$. They are defined by

$$\psi_{a,b}(t) = \frac{1}{\sqrt{|a|}} \psi \frac{t-b}{a}, , a,b \in R , a \neq 0 \quad (1)$$

The parameter a is the scaling parameter or scale, and it measures the degree of compression. The parameter b is the translation parameter which determines the time location of the wavelet. Wavelet transforms provide a framework in which a signal is decomposed, with each level corresponding to a coarser resolution or lower-frequency band and higher-frequency bands.

Wavelet-based image fusion model

The most common widely used transform for image fusion at multi modal is Wavelet Transform since it minimizes structural distortions. The wavelet based image fusion model first involves the DWT of the input images. The decomposition is performed on each input image separately to obtain their wavelet sub images. A set of fusion rules are then applied to the sub image coefficients and these coefficient values are modified based on the desired specifications. This results in a new set of coefficients representing the combination of desired properties from each input image. The inverse wavelet transform is applied to these new coefficients. This reconstruction results in the fused image.

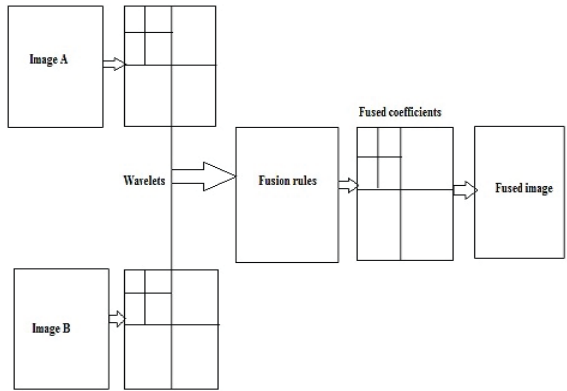

Fig 1. Wavelet Based Image Fusion.

$$I(x, y) = w^{-1}(\Phi(w(I_1(x, y)), w(I_2(x, y)))) \quad (2)$$

where $I_1(x, y)$ and $I_2(x, y)$ are images to be fused, the decomposed low frequency sub images of $I_1(x, y)$ and $I_2(x, y)$ be respectively $lI_{1j}(x, y)$ and $lI_{2j}(x, y)$ ( J is the parameter of resolution) and the decomposed high frequency sub images of $I_1(x,y)$ and $I_2(x,y)$ are $hI_{1j}^k(x, y)$ and $hI_{2j}^k(x, y)$. (j is the parameter of resolution and j=1,2,3….J for every j, k=1,2,3..)[12]. Then the inverse wavelet transform $w^{-1}$ is computed and the fused image $I(x, y)$ is reconstructed. There are several wavelet fusion rules that can be used for the selection of the wavelet coefficients from the wavelet transforms of the images to be fused. The most frequently used rule is the maximum frequency rule which selects the coefficients that have the maximum absolute values [11]. Wavelet transform suffers from lack of shift invariance & poor directionality and these disadvantages are overcome by Stationary Wavelet Transform and Dual Tree Wavelet Transform.

Discrete Wavelet Transform

The discrete wavelet transform (DWT) is a spatial-frequency decomposition that provides a flexible mutliresolution analysis of an image. In one dimension the aim of the wavelet transform is to represent the signal as a superposition of wavelets. If a discrete signal is represented by f (t), its wavelet decomposition is then

$$f(t) = \sum_{m,n} c_{m,n} \psi_{m,n}(t) \quad (3)$$

Where $\psi_{m,n}(t)$ is the dilated and/or translated vision of the mother wavelet $\psi$ given by the equation

$$\psi_{m,n}(t) = 2^{-m/2} \psi[2^{-m}t - n] \quad (4)$$

where m and n are integers. This ensures that the signal is decomposed into normalized wavelets at octave scales. When a 1-D DWT is first performed on the rows and then columns of the data by separately filtering and down sampling. This results in one set of approximation coefficients and three sets of detail coefficients, which represent the horizontal, vertical and diagonal directions of the image, respectively. The 2-D DWT produces three band-pass sub images at each level, which are corresponding to LH, HH, HL, and oriented at angles of 0°, ± 45°, 90°.

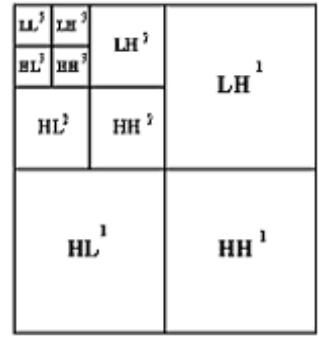

Fig 2: 2-D discrete wavelets transform

In the language of filter theory, these four sub-images correspond to the outputs of low–low (LL), low–high (LH), high–low (HL), and high–high (HH) bands. By recursively applying the same scheme to the LL sub-band a multi-resolution decomposition with a desires level can then be achieved. The reconstructed tasks has been hampered by two disadvantages: *lack of shift invariance*, which means that small shifts in the input signal can cause major variations in the distribution of energy between DWT coefficients at different scales; *poor directional selectivity* for diagonal features, because the wavelet filters are separable and real.

## Fast Wavelet Transform

The Fast wavelet transform is computationally efficient implementation of the DWT. The DWT matrix is not sparse in general, so we face the same complexity issues in DWT as same as faced for the discrete Fourier transform (7). Then solve it as same as the FFT, by factoring the DWT into a product of a few sparse matrices using self-similarity properties. The result is an algorithm that requires only order *n* operations to transform an *n*-sample vector. This is the "fast" DWT of Mallat and Daubechies. It resembles the two band sub band coding scheme and also called Mallat's herringbone algorithm.

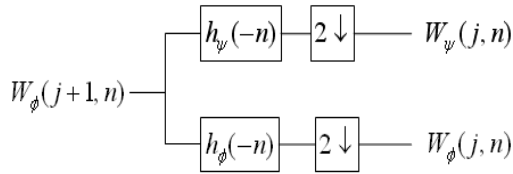

Fig 3: An FWT analysis filter bank

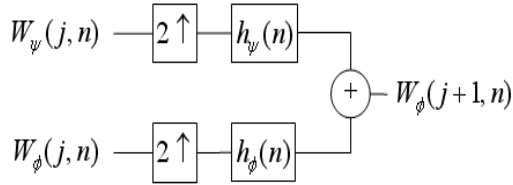

Fig 4: An FWT$^{-1}$ synthesis filter bank.

$$W_\psi(j,k) = h_\psi(-n) * W_\phi(j+1,n)\big|_{n=2k, k \geq 0}$$

$$W_\phi(j,k) = h_\phi(-n) * W_\phi(j+1,n)\big|_{n=2k, k \geq 0}$$

By sub band coding theorem, perfect reconstruction for two-band orthonormal filters requires $g_i(n) = h_i(-n)$ for $i = \{0, 1\}$. That is, the synthesis and analysis filters must be time-reversed versions of one another. Since the FWT analysis filter are $h_0(n) = h_\Phi(-n)$ and $h_1(n) = h_\Psi(-n)$, the required FWT$^{-1}$ synthesis filters are $g_0(n) = h_0(-n) = h_\Phi(n)$ and $g_1(n) = h_1(-n) = h_\Psi(n)$.

## Wavelet Packet Transform

The wavelet transform is actually a subset of a far more versatile transform, the wavelet packet transform. Wavelet packets are particular linear combinations of wavelets [8]. They form bases which retain many of the orthogonality, smoothness, and localization properties of their parent wavelets. The coefficients in the linear combinations are computed by a recursive algorithm making each newly computed wavelet packet coefficient sequence the root of its own analysis tree. The families of orthonormal basis associated with this binary-tree (WP-tree) are known as 'wavelet packets'. The complexity of computation for WP is O (n log n). WPT is used for Compression and de-noising.

### III. STATIONARY WAVELET TRANSFORM

The Stationary wavelet transforms (SWT) is not time invariant transform. The SWT has a similar tree structured implementation without any decimation (sub sampling) step. SWT has equal length wavelet coefficients at each level. The computational complexity of SWT is O (n)$^2$. The redundant representation makes SWT shift-invariant and suitable for applications such as edge detection, denoising and data fusion. The translation-invariance is achieved by removing the down samplers and up samplers in the DWT and up sampling the filter coefficients by a factor of 2(j − 1) in the jth level of the algorithm [7]. The SWT is an inherently redundant scheme as the output of each level of SWT contains the same number of samples as the input – so for a decomposition of N levels there is a redundancy of N in the wavelet coefficients. This algorithm is more famously known as "algorithme à trous" in French what refers to inserting zeros in the filters. It was introduced by Holdschneider. When stationary wavelet is applied to the image it undergoes decomposition and reconstruction. Decomposition of an image results in approximation (low-low frequency sub-bands) and detailed coefficients (low-high, high-low and high-high frequency sub-bands).

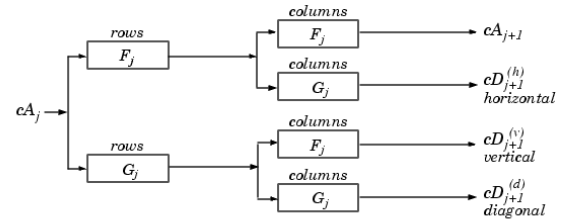

Fig 5. 2D Stationary Wavelet Transform.

Shift-variance is caused by the decimation process, and can be resolved by using the un-decimated algorithm. Let us recall that the DWT basic computational step is a convolution followed by decimation. The decimation retains even indexed elements. But the decimation could be carried out by choosing odd indexed elements instead of even indexed elements. This choice concerns every step of the decomposition process, so at every level we chose odd or even. If we perform all the different possible decompositions of the original signal, we have 2J different decompositions, for a given maximum level J. Let us denote by j = 1 or 0 the choice of odd or even indexed elements at step j. Every decomposition is labeled by a sequence of 0's and 1's: = 1, J. This transform is called the decimated DWT. It is possible to calculate all the decimated DWT for a given signal of length N, by computing the approximation and detail coefficients for every possible sequence. The SWT algorithm is very simple and is close to the DWT one. More precisely, for level 1, all the decimated DWT for a given signal can be obtained by convolving the signal with the appropriate filters as in the DWT case but without down sampling. Then the approximation and detail coefficients at level 1 are both of size N, which is the signal length. The general step j convolves the approximation coefficients at level j-1, with up sampled versions of the appropriate original filters, to produce the approximation and detail coefficients at level j. This can be visualized in the following figure 5.

## IV. INTEGER LIFTING WAVELET TRANSFORM

The conventional convolution-based implementation of the discrete wavelet transform has high computational and memory requirements. Recently, the lifting-based implementation of the discrete wavelet transform has been proposed to overcome these drawbacks and named as Lifting Wavelet Transform (LWT). LWT is also called second generation wavelet transform. In spatial domain, the realization process of lifting wavelet transform mainly divided into three steps: split, prediction and update. The original input signal is $f_k$. It is transformed into signal of high pass $h_k$ and low pass signal $l_k$. In the split step, the original signal is split into two non-overlap subsets, namely even sequence and odd sequence. In the prediction step, even sequences are used to predict odd sequences. The prediction error forms the corresponding high-pass sub band. In the update step, an approximation sub band is obtained by updating even sequences with the scaled high-sub band samples, which forms a low-pass sub band. Backward transform is easy to find and has the same complexity as the forward transform. The two-step lifting transform can be generally described as

$h_k[x] = f_{2k+1}[x] + \sum_i P_i f_{2(k-i)}[x]$ (5)

$l_k[x] = f_{2k}[x] + \sum_j u_j h_{k-j}[x]$ (6)

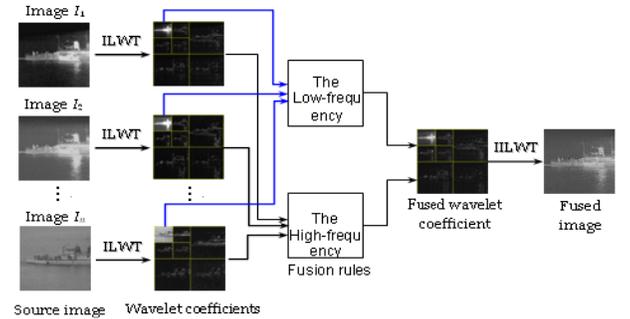
Fig 6: lifting wavelet transform based image fusion

where $f_k[x]$ is the sequence of input data to be processed, $h_k$ and $l_k$ are resulting high-pass and low-pass sequences respectively, $P_i$ and $u_j$ are prediction and update coefficients of filters respectively.

## V. DUAL TREE WAVELET TRANSFORM

The dual-tree complex wavelet transform is a recent enhancement to the discrete wavelet transform [10]. The transform was first proposed by Kingsbury [9] in order to mitigate two main disadvantages, namely, the lack of shift invariance and poor directional selectivity, of the discrete wavelet transform (DWT). There are two versions of the 2D DTWT transform namely Dual Tree Discrete Wavelet Transform (DTDWT) which is 2-times expansive, and Dual Tree Complex Wavelet Transform (DTCWT) which is 4-times expansive. The dual-tree CWT employs two real DWTs; the first DWT gives the real part of the transform while the second DWT gives the imaginary part. The two real wavelet transforms use two different sets of filters, with each satisfying the perfect reconstruction conditions. The two sets of filters are jointly designed so that the overall transform is approximately analytic. The inverse of the dual-tree CWT is as simple as the forward transform. To invert the transform, the real part and the imaginary part are each inverted—the inverse of each of the two real DWTs are used—to obtain two real signals[10]. These two real signals are then averaged to obtain the final output. The properties of the DT-CWT can be summarized as
    a) approximate shift invariance;
    b) good directional selectivity in 2 dimensions;
    c) phase information;
    d) perfect reconstruction using short linear-phase filters;

e) limited redundancy, independent of the number of scales, 2 : 1 for 1D (2m : 1 for mD);

f) efficient order-N computation—only twice the simple DWT for 1D (2m times for mD).

It has the ability to differentiate positive and negative frequencies and produces six subbands oriented in ±15°, ± 45°, ±75°. Fig. 2 shows the impulse responses of the dual-tree complex wavelets. It is evident that the transform is selective in 6 directions in all of the scales except the first.

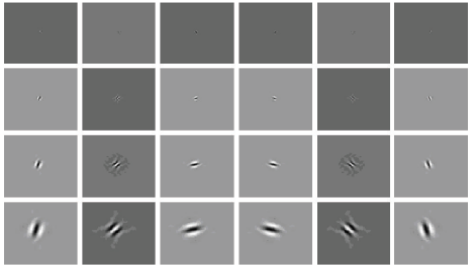

Fig 7. Impulse response of dual-tree complex wavelets at 4 levels and 6 directions.

Unfortunately the odd/even filter approach suffers from certain problems: a) the sub-sampling structure is not symmetrical; b) the two trees have slightly different responses; and c) the filter sets must be bi-orthogonal. To overcome all problems above, Kingsbury proposed a Q-shift dual tree CWT.

## VI. Q-SHIFT DUAL TREE COMPLEX WAVELET TRANSFORM

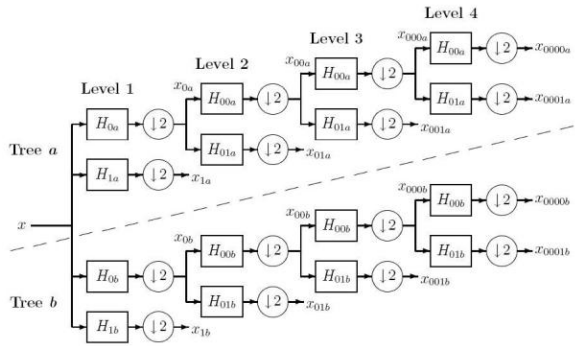

Fig 8. The Q-shift dual tree structure

Q-shift dual tree wavelet is a more recent form of dual tree wavelet. Q-shift dual tree is shown in figure 1. There are two sets of filters used, the filter at level 1, and the filters at all higher levels. as in fig 8, in which all the filters beyond level 1 are even length, but they are no longer strictly linear phase. Instead they are designed to have a group delay of approximately ¼ samples (q). The required delay difference of ½ samples (2q) is then achieved by using the time reverse of the tree a filters in tree b. This leads to a more symmetric sub-sampling structure, but which preserves the key advantages of DT-CWT that are approximate shift invariance and good directional selectivity. Therefore we decompose the input images with Q-shift CWT. There are a number of choices of possible filter combinations. We have chosen to use the (13-19)-tap near-orthogonal filters at level 1 together with the 14-tap Q-shift filters at levels ≥ 2 [19]. The Q-shift transform retains the good shift invariance and directionality properties of the original while also improving the sampling structure. When we talk about the complex wavelet transform we shall always be referring to this Q-shift version unless explicitly stated otherwise. We will often refer to this transform by the initials DT-CWT. The DT-CWT achieves a very low noise gain and so will give robust reconstructions. The Q-shift tree has a lower noise gain than the original dual tree. This is because of the better balanced filters in the Q-shift version. The single tree complex wavelets which is likely to make the wavelets useless. So, we choose Q-shift dual tree complex wavelet transform.

## VII. QUALITY EVALUATION

The quality of the fused image is evaluated by four benchmarks: Entropy (E), Root Mean Square Error (RMSE), Peak Signal to Noise Ratio (PSNR), Quality Index (QI)[3] and Normalized Weighted Performance Metric (NWPM)[4] and Mutual Information[MI]. Consider R as the source image and F the fused image, both of size M×N. F (i, j) is the grey value of pixel at the position (i, j).

1) Entropy (E)

Entropy is an index to evaluate the information quantity contained in an image. If the value of entropy becomes higher after fusing, it indicates that the information increases and the fusion performance are improved.

$$E = -\sum_{i=0}^{L-1} p_i \, log_2 p_i$$

Entropy is defined as where L is the total of grey levels, p= {p0, p1…pL-1} is the probability distribution of each level [14].

2) Peak-to-Peak Signal-to-Noise Ratio (PSNR)

PSNR is the ratio between the maximum possible power of a signal and the power of corrupting noise that affects the fidelity of its representation. The PSNR measure is given by [14]

PSNR= $10 log\ 10\ (255)^2 / (RMSE)^2$ (db)

3) Root Mean Square error

The root Mean Square error is defined as follows:

$$RMSE = \sqrt{\frac{\sum_{i=1}^{M}\sum_{j=1}^{N}[R(i,j)-F(i,j)]^2}{M \times N}}$$

The RMSE is used to measure the difference between the source image and the fused image; the smaller the value of RMSE and the smaller the difference, the better the fusion performance.

4) Image Quality Index (IQI)

IQI was introduced by Wang and bovik [13]. Given two images x and y. It measures the similarity between two images and its values ranges from -1 to 1. IQI is equal to 1 if both the images are identical. IQI measure is given by

$$IQI = 4\sigma_{xy}xy/(x^2+y^2)(\sigma_x^2+\sigma_y^2)$$

where $\sigma_x^2$, $\sigma_y^2$, $\sigma_{xy}$ denotes the variance of x, y and covariance of x and y respectively.

5) Standard deviation (SD)

Standard deviation is shown as follows

$$SD = \sqrt{\frac{1}{M \times N}\sum_{m=1}^{M}\sum_{n=1}^{N}(F(m,n)-MEAN)^2}$$

Where MEAN is the average denoted by

$$MEAN = \frac{1}{M \times N}\sum_{m=1}^{M}\sum_{n=1}^{N}|F(m,n)|$$

## VIII. EXPERIMENTAL RESULTS

The experimental results of the five mentioned fusion methods are displayed in fig.9 (c)-(g), respectively. The pair of source images to be fused is assumed to be registered spatially. The images are wavelet transformed using the same wavelet, and transformed to the same number of levels. For taking the wavelet transform of the two images, readily available MATLAB routines are taken. A fused wavelet transform is created by taking pixels from that wavelet transform that shows greater activity at the level. The inverse wavelet transform is the fused image with clear focus on the whole image. For the above mentioned method, image fusion is performed using DWT, SWT, ILWT DTCWT and Q-shift DT-CWT; their performance is measured in terms of Entropy, Peak Signal to Noise Ratio, Root Mean Square Errors, Image Quality Index & Standard deviation and tabulated in table1.

## IX. CONCLUSIONS

The aim of this paper has been to compare all levels of fusion of multi modal images using DWT, SWT, ILWT, DTCWT and Q-shift DTCWT in terms of various performance measures. The Q-shift DT-CWT is a valuable enhancement of the traditional real wavelet transform that is nearly shifting invariant and, in higher dimensions, directionally selective. The developed Q-shift DTCWT fusion technique provides computationally efficient and better qualitative and quantitative results. The Q-shift DTCWT method is used to retain edge information without significant ringing artifacts. It has the further advantages that the phase information is available for analysis

## X REFERENCES

1. Tian Hui, Wang Binbin, "*Discussion and Analyze on Image Fusion Technology*", 2009.
2. Strang, G. Wavelets and Dilation Equations: A brief introduction. *SIAM Review,* 31: 614-627, 1989.
3. Walnut, D.F. *An Introduction to Wavelet Analysis*. Birkhäuser, Boston, 2001.
4. Wells, R.O. Parametrizing Smooth Compactly Supported Wavelets.*Transform American Mathematical Society,* 338(2): 919-931, 1993.
5. Debnath, L. *Wavelet Transformation and their Applications*. Birkhäuser Boston, 2002.
6. M. I. Smith, J. P. Heather, "*Review of Image Fusion Technology in 2005*," Proceedings of the SPIE, Volume 5782, pp. 29-45, 2005.
7. G. Pajares, J. M. D. L. Cruz, "*A wavelet-based image fusion tutorial*, "Pattern Recognition, vol. 37, no. 9, pp. 1855–1872, 2004. Kanagaraj.
8. M.A. Cody, \The Wavelet Packet Transform," *Dr. Dobb's Journal,* Vol 19, Apr. 1994, pp. 44- 46, 50-54.
9. N. G. Kingsbury, "*The dual-tree complex wavelet transform:* A new technique for shift invariance and directional filters," in *Proc. 8th IEEE DSP Workshop*, UT, Aug. 1998.
10. I. W. Selesnick, R. G. Baraniuk, and N. G. Kingsbury, "The dual-tree complex wavelet transform – *A coherent framework for multiscale signal and image processing*," *IEEE Signal Process. Mag.*, vol. 6, pp. 123–151, Nov. 2005.
11. Udomhunsakul, S. and P. Wongsita, "Feature extraction in medical MRI images", proceeding of 2004 *IEEE conference* on cybernetics and intelligent Systems, Vol. 1,340-344, Dec. 2004.
12. Paul Hill, Nishan Canagarajah and Dave Bull "*Image Fusion using Complex Wavelets*" Dept. of Electrical and Electronic Engineering The University of Bristol Bristol, BS5 lUB, UK, 2002
13. Zhou Wang and Alan C. Bovik, "A Universal Image Quality Index", IEEE Signal Processing Letters, Vol. 9, No.3, pp. 81-84, March, 2.
14. Hossam EI_Din Moustafa, Sameh Rehan, "Applying Image Fusion Techniques for Detection of Hepatic Lesions and Acute Intra-

Cerebral Hemorrhage" Communications and Electronics Engineering Department, Faculty of Engineering, Mansoura University, Mansoura, EGYPT 35516.

Table 1. Evaluation results of the four different fusion techniques

| Bench marks | DWT | SWT | ILWT | DT-CWT | Q-shift DT-CWT |
|---|---|---|---|---|---|
| EN | 5.2187 | 6.9234 | 6.9345 | 6.9441 | 6.9642 |
| PSNR | 35.4296 | 37.0541 | 37.3868 | 37.9704 | 37.9898 |
| RMSE | 4.3158 | 3.4491 | 3.3456 | 3.2212 | 3.1012 |
| IQI | 0.9843 | 0.9851 | 0.9866 | 0.9873 | 0.9892 |
| SD | 39.455 | 40.364 | 40.689 | 40.973 | 40.986 |

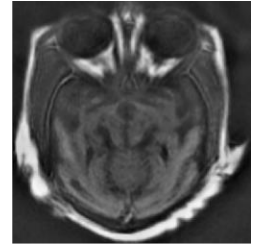

g)

Fig 9: a. CT image b. MRI image c. Fused Image using DWT d. Fused Image using SWT e. Fused Image using ILWT f. Fused Image using DT-CWT g. Fused Image using Q-shift DT-CWT

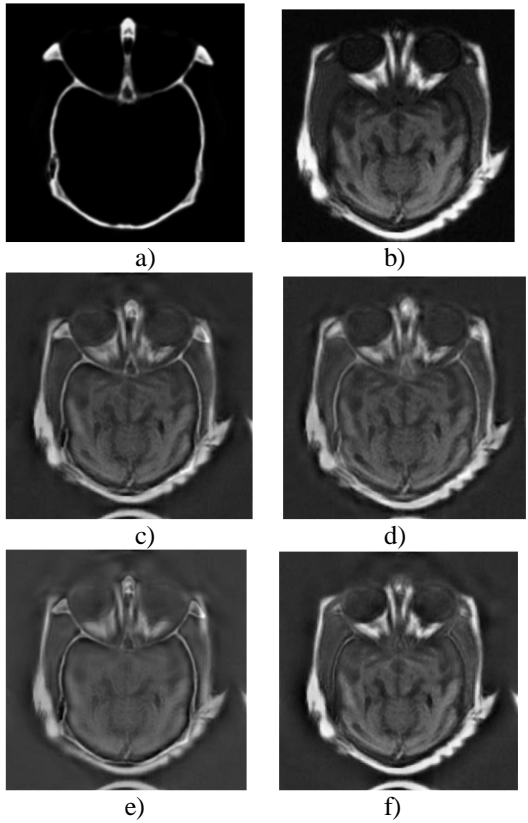

a)     b)

c)     d)

e)     f)